\documentclass[useAMS,usenatbib, usegraphicx]{mn2e}
 \usepackage[dvips]{color}

\newcommand{\kms}{\mbox{km}\,\mbox{s}^{-1}}
\newcommand{\target}{KPD\,1946+4340}

\title[{\em Kepler} observations of the beaming binary \target]{{\em Kepler} observations of the beaming binary \target}

\author[S. Bloemen et al.]{S. Bloemen$^{1}$\thanks{E-mail:
steven.bloemen@ster.kuleuven.be}, T. R. Marsh$^{2}$, R. H. \O stensen$^{1}$, S. Charpinet$^{3}$, G. Fontaine$^{4}$,  \newauthor P. Degroote$^{1}$, U. Heber$^{5}$, S. D. Kawaler$^{6}$,
C. Aerts$^{1,7}$, E. M. Green$^{8}$, J. Telting$^{9}$, \newauthor  P. Brassard$^{4}$, B. T. G\"ansicke$^{2}$, G. Handler$^{10}$, D. W. Kurtz$^{11}$, R. Silvotti$^{12}$, 
\newauthor V. Van Grootel$^{3}$, J. E. Lindberg$^{8,13}$, T. Pursimo$^{8}$, P. A. Wilson$^{8,14}$, \newauthor R. L. Gilliland$^{15}$, H. Kjeldsen$^{16}$, 
J. Christensen-Dalsgaard$^{16}$,  W. J. Borucki$^{17}$,\newauthor D. Koch$^{17}$, J. M. Jenkins$^{18}$, T. C. Klaus$^{19}$\\ 
$^{1}$Instituut voor Sterrenkunde, Katholieke Universiteit Leuven, Celestijnenlaan 200D, B-3001 Leuven, Belgium\\
$^{2}$Department of Physics, University of Warwick, Coventry CV4 7AL, UK\\
$^{3}$Laboratoire d'Astrophysique de Toulouse-Tarbes, Universit\'e de Toulouse, CNRS, 14 Av. E. Belin, 31400 Toulouse, France\\
$^{4}$D\'epartement de Physique, Universit\'e de Montr\'eal, C.P. 6128, Succ. Centre-Ville, Montr\'eal, Qu\'ebec H3C 3J7, Canada\\
$^{5}$Dr.~Remeis-Sternwarte \& ECAP Astronomisches Institut, Univ.~Erlangen-N\"urnberg, Sternwartstr. 7, 96049 Bamberg, Germany\\
$^{6}$Department of Physics \& Astronomy, Iowa State University, Ames, IA 50011, USA\\
$^{7}$Department of Astrophysics, IMAPP, Radboud University Nijmegen, PO Box 9010, NL-6500 GL Nijmegen, the Netherlands\\
$^{8}$Steward Observatory, University of Arizona, 933 North Cherry Avenue, Tucson, AZ 85721, USA\\
$^{9}$Nordic Optical Telescope, 38700 Santa Cruz de La Palma, Spain\\
$^{10}$Institut f\"ur Astronomie, Universit\"at Wien, T\"urkenschanzstrasse 17, 1180 Wien, Austria\\
$^{11}$Jeremiah Horrocks Institute of Astrophysics, University of Central Lancashire, PR1 2HE, UK\\
$^{12}$INAF-Osservatorio Astronomico di Torino, Strada dell'Osservatorio 20, 10025 Pino Torinese, Italy\\
$^{13}$Centre for Star and Planet Formation, Natural History Museum of
Denmark, University of Copenhagen, \O ster Voldgade 5-7, \\ \ \ \ DK-1350
Copenhagen, Denmark\\
$^{14}$Institute of Theoretical Astrophysics, University of Oslo, P.O. Box 1029 Blindern, N-0315 Oslo, Norway\\
$^{15}$Space Telescope Science Institute, 3700 San Martin Drive, Baltimore, MD 21218, USA\\
$^{16}$Department of Physics and Astronomy, Aarhus University, DK-8000 Aarhus C, Denmark\\
$^{17}$NASA Ames Research Center, MS 244-30, Moffett Field, CA 94035, USA\\
$^{18}$SETI Institute/NASA Ames Research Center, MS 244-30, Moffett Field, CA 94035, USA\\
$^{19}$Orbital Sciences Corp., NASA Ames Research Center, MS 244-30, Moffett Field, CA 94035, USA}
\begin{document}
\date{Accepted 2010 August 18. Received 2010 August 17; in original form 2010 June 5}

\pagerange{\pageref{firstpage}--\pageref{lastpage}} \pubyear{2010}

\maketitle

\label{firstpage}

\clearpage
\begin{abstract}
The {\em Kepler Mission} has acquired 33.5\,d of continuous one-minute
photometry of \target, a short-period binary system that consists of a subdwarf B star (sdB) and a white dwarf.
In the light curve, eclipses are clearly seen, with the deepest
occurring when the compact white dwarf crosses the disc of the sdB (0.4 per cent) and
the more shallow ones (0.1 per cent) when the sdB eclipses the white dwarf. As expected,
the sdB is deformed by the gravitational field of the white dwarf, which produces
an ellipsoidal modulation of the light curve.  Spectacularly,
a very strong Doppler beaming (also known as Doppler boosting)
effect is also clearly evident at the 0.1 per cent level. This originates
from the sdB's orbital velocity, which we measure to be $164.0\pm1.9\,\kms$ from supporting spectroscopy.
We present light curve models that account for all these effects, as well as
gravitational lensing, which decreases the apparent radius of the white dwarf
by about 6 per cent when it eclipses the sdB.
We derive system parameters and uncertainties from the light curve
using Markov Chain Monte Carlo simulations. Adopting a theoretical white
dwarf mass-radius relation, the mass of the subdwarf is found to be
$0.47\pm0.03\,$M$_\odot$ and the mass of the white dwarf $0.59\pm0.02\,$M$_\odot$. The effective
temperature of the white dwarf is $15\,900\pm300\,$K.
With a spectroscopic effective temperature of $T_{\rm{eff}}=34\,730\pm250\,$K and a surface gravity
of $\log g=5.43\pm0.04$, the subdwarf has most likely exhausted its core helium, and is
in a shell He burning stage.

The detection of Doppler beaming in Kepler light curves potentially
allows one to measure radial velocities without the need of
spectroscopic data.
For the first time, a photometrically observed Doppler beaming amplitude
is compared to a spectroscopically established value.
The sdB's radial velocity amplitude derived from the photometry ($168\pm4\,\kms$) is in perfect agreement with the spectroscopic value. After subtracting our best model for the orbital effects, we searched the residuals for stellar oscillations but did not find any significant pulsation frequencies.
\end{abstract}



\section[]{Introduction}\label{sec_intro}

Subdwarf B stars are mostly assumed to be extreme horizontal
branch stars, i.e., core helium burning stars with a thin
inert hydrogen envelope \citep{Heber1986, SafferBergeron1994}.
In order to reach such high temperatures and surface gravities,
the progenitor must have lost almost its entire hydrogen envelope.
The majority of sdBs is expected to have lost its envelope via binary interaction channels,
as elaborated by \citet{HanPodsiadlowski2002, HanPodsiadlowski2003}. Our
target, \target\ (KIC 7975824), is a subdwarf B star (sdB) with a white dwarf (WD) companion in a
0.403739(8) day orbit (Morales-Rueda et al. 2003), which identifies
the theoretical formation channel for this system as
the second common-envelope ejection channel of
Han et al. (2002, 2003). In this scenario the white dwarf is
engulfed by the sdB progenitor as it ascends the first
giant branch. The white dwarf will deposit its angular momentum in the
atmosphere of the giant and spin up the envelope until it
is ejected. There are two subchannels to this scenario,
depending on the initial mass of the progenitor.
If sufficiently massive, it will ignite helium
non-degeneratively, and the resulting extended horizontal branch (EHB) star will have a
mass of $\sim 0.35\,$M$_\odot$. The more common scenario, starting
with a roughly solar-mass giant, produces an EHB star with
a mass that must be very close to the helium flash mass of
 $0.47\,$M$_\odot$. A third possibility occurs when the white dwarf companion
ejects the envelope before the core has attained sufficient
mass to ignite helium. In this case the remaining core will
evolve directly to the white dwarf cooling track. On its way it crosses
the domain of the EHB stars, but without helium ignition
the period in which it appears as an sdB star is brief,
making this channel a very small contributor to the sdB
population.
For a recent extensive review on hot subdwarf stars, their
evolution and observed properties, see \citet{Heber2009}.

The exact physical details involved in common-envelope ejection are not well
understood. This uncertainty is commonly embodied in the
efficiency parameter $\alpha$, which denotes the amount of
orbital energy used to eject the envelope \citep[see e.g.][]{de-Kool1990, HuNelemans2007}.
Eclipsing subdwarf binaries could help constrain the permitted
values of $\alpha$, but studies have hitherto been hampered
by the fact that both sdB+WD and sdB+M-dwarf binaries have virtually
invisible companions and are therefore single lined, leaving the masses
indeterminate. Firmly establishing the parameters of both components
of a post-CE system therefore has substantial implications not just
for confirming that our formation scenarios are correct, but also
in order to tune future binary population synthesis studies by
confining the $\alpha$ parameter.

The target studied here, \target, is an sdB star discovered by
the {\em Kitt Peak Downes} survey \citep{Downes1986}.  \target\ has a $V$-band magnitude of $14.284\pm0.027$ \citep{AllardWesemael1994}, a $y$-magnitude of $14.299\pm0.002$ \citep{WesemaelFontaine1992} and a {\em Kp} ({\em Kepler}) magnitude of 14.655. 
The star was included
in the radial velocity survey of \citet{Morales-RuedaMaxted2003}, who found
the target to be a spectroscopic binary with a period of 0.403739(8)\,d and
a velocity amplitude $K_1=167\pm2\,\kms$.
They also concluded that the sdB primary should be in a post-EHB
stage of evolution, due to its relatively low surface gravity placing
it above the canonical EHB in the HR diagram. This implies that the sdB exhausted all available helium in its core and is now in a shell helium burning stage. Assuming the sdB mass to be $0.5\,$M$_\odot$, they found a minimum mass of
$0.628\,$M$_\odot$ for the companion. 

In this paper we present the first light curve of \target\ obtained
from space. The target was observed for 33.5\,d by the {\em Kepler Mission},
and the light curve reveals sufficient low level features to permit
purely photometric measurements of velocities, radii and masses of both
components.  A review of the {\em Kepler Mission} and its first results is given in \citet{KochBorucki2010}.

We combine the {\em Kepler} photometry with new and old spectroscopic
measurements, use light curve modelling to estimate the system
parameters and Markov Chain Monte Carlo (MCMC) simulations to establish
the uncertainties.
The relativistic Doppler beaming effect is clearly detected in the
light curve, and can be used to determine the orbital velocity of
the primary.  This effect, which is also known
as Doppler boosting, was recently noted in a {\em Kepler} light
curve of KOI-74 by van Kerkwijk et al.~(2010).
We present the first comparison of a radial velocity amplitude as
derived from the amplitude of the Doppler beaming to the
spectroscopically determined value.
We also use the spectroscopic data to provide a revised ephemeris,
as well as to determine the effective temperature,
surface gravity and helium fraction of the atmosphere.
After detrending the {\em Kepler} light curve with our best model
for the orbital effects, we search the residuals for stellar oscillations.


\section[]{Observations, radial velocities and updated ephemeris} \label{sec_obs} 
\begin{figure*}
\includegraphics[width=180mm]{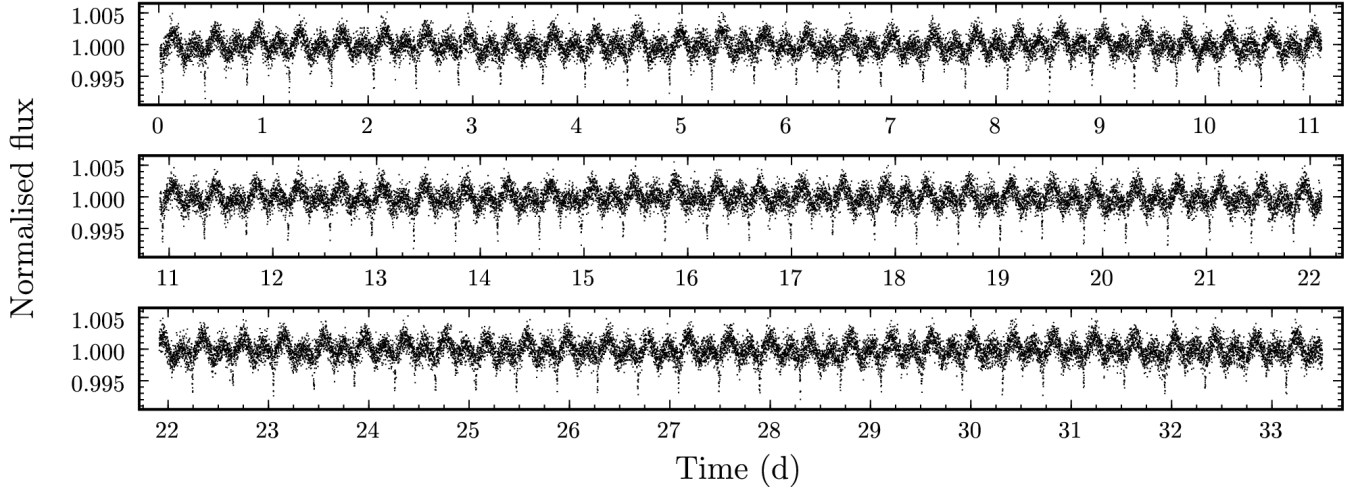}
 \caption{{\em Kepler} light curve of \target\ after detrending and removing outliers.}
  \label{FIG_LC}
\end{figure*}

We used 33.5\,d of Q1 short cadence {\em Kepler} data with a time resolution of 59\,s. A
review of the characteristics of the first short cadence datasets is presented
by \citet{GillilandJenkins2010}. The data were delivered to us through the
KASOC ({\em Kepler} Asteroseismic Science Operations Center) website\footnote{http://kasoc.phys.au.dk/kasoc/}. The level of contamination of the fluxes by other stars is poorly known. We used the raw fluxes and assumed zero contamination, which is justified by the absence of other significantly bright sources within ten arcsec of  \target.  We applied a
barycentric correction to the {\em Kepler} timings and converted them from UTC to
barycentric dynamical time (TDB). The raw data show a $\sim 2$ per cent downward trend over the 33.5\,d which we assume is instrumental in
origin. We removed this variation by fitting and dividing out a spline
function. Out of the original dataset of 49170, we rejected 54 points because of a bad quality flag. Then, after initial light curve model fits to be described below, we
rejected another 87 points because they differed by more than $3.5\, \sigma$ from our model. The full light curve we used for our analysis is shown on Fig.~\ref{FIG_LC}. The time span of the dataset is
BMJD(TDB)\footnote{BMJD(TDB) refers to Barycentric-corrected Modified Julian Date on the Barycentric Dynamical Timescale.} 54\,964.00314 to 54\,997.49381. 

The high signal-to-noise spectra from \citet{GreenFontaine2008} were used to derive $T_{\rm{eff}}$ and $\log g$; see Section \ref{sec_spec} for details.

\begin{figure}
\includegraphics[width=84mm]{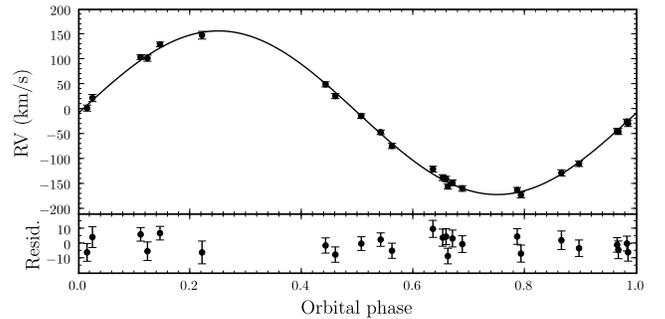}
 \caption{Radial velocity curve of the sdB in \target. Both our new radial velocity measurements and the ones from \citet{Morales-RuedaMaxted2003} are shown, folded on the orbital period. The error bars of the datapoints show the uncertainties after adding $3.6\,\kms$ and $4.1\,\kms$ in quadrature to the values of \citet{Morales-RuedaMaxted2003} and our new data respectively. We find a radial velocity amplitude of $K_1=164.0\pm1.9\,\kms$. }
  \label{FIG_RV}
\end{figure}

To refine the orbital period determination of \citet{Morales-RuedaMaxted2003},
spectra were collected with the {2.56-m} Nordic Optical Telescope (NOT) on
2009 December 5, 9 and 10. Using the ALFOSC spectrograph and a 0.5 arcsec slit, we obtained eleven spectra with exposure times of 300\,s and a of resolution $R\sim2000$, covering a wavelength region of $3500-5060$\AA. We measured the radial velocities using
multi-Gaussian fits \citep{Morales-RuedaMaxted2003}; the values are listed in
Table \ref{tab_RV}. We first fitted the data of
\citet{Morales-RuedaMaxted2003} and the new data separately to determine their
RMS scatter. From these fits we found that it was necessary to add $3.6$ and
$4.1\,\kms$ in quadrature to the uncertainties of the two datasets to deliver
a unit $\chi^2$ per degree of freedom; these values probably reflect
systematic errors due to incomplete filling of the slit. We then carried out a
least-squares sinusoidal fit to the combined dataset finding a best fit of
$\chi^2 = 21.6$ (25 points). The next best alias had $\chi^2 = 39$, and so we
consider our best alias to be the correct one. This gave the following
spectroscopic ephemeris:
\[\mbox{BMJD(TDB) }= 53\,652.84813(62) + 0.40375026(16) E,\]
marking the times when the sdB is closest to Earth.  The corresponding radial
velocity amplitude was $K_1=164.0\pm1.9\,\kms$. The WD's spectrum will slightly reduce the observed velocity amplitude of the sdB. We estimate that this effect is less than $1\, \kms$. The radial velocity measurements and the fit are shown in Fig.~\ref{FIG_RV}. From fitting light curve
models to the {\em Kepler} photometry (see Section \ref{sec_lc_mod}) we
derived the following photometric ephemeris:
\[ \mbox{BMJD(TDB) }=  54\,979.975296(25) + 0.40375000(96) E.\]
The two independent periods agree to within their uncertainties. The 9\,yr baseline of the
spectroscopic ephemeris gives a more precise value and henceforth we fix it at
this value in our lightcurve models. The cycle count between the spectroscopic and
photometric ephemerides (using the spectroscopic period) is
$3287.0001\pm0.0015$, an integer to within the
uncertainties. The {\em Kepler}-based zeropoint is the more precise one and is
therefore retained as a free parameter in our models. 

\begin{table}
 \caption{Radial velocities of the sdB in \target\ determined from NOT spectroscopy.}
 \label{tab_RV}
 \begin{center}
 \begin{tabular}{cr}
  \hline
BMJD (TDB) & RV ($\kms$)\\
\hline

55170.81118&$ -139.8 \pm 3.5$\\
55170.86268&$ -162.0 \pm 3.3$\\
55170.90743&$ -109.2 \pm 3.8$\\
55170.95896&$ 22.5 \pm 5.7$\\
55174.80952&$ -73.5 \pm 3.0$\\
55174.84999&$ -154.7 \pm 3.3$\\
55174.90291&$ -171.9 \pm 4.0$\\
55174.93220&$ -127.6 \pm 4.7$\\
55175.80006&$ 2.3 \pm 4.4$\\
55175.84379&$ 102.4 \pm 4.7$\\
55175.88328&$ 148.3 \pm 6.5$\\

\hline
  \end{tabular} \end{center}
\end{table}

\section[]{Light curve analysis} \label{sec_lc} 
The {\em Kepler} light curve we analyse in this paper (Fig.~\ref{FIG_LC}) reveals that \target\ is an eclipsing binary. We graphically determined the eclipse depths and durations. The eclipses of the WD by the sdB are $0.13\pm0.03$ per cent deep and the eclipses of the sdB by the white dwarf $0.38\pm 0.03$ per cent. The duration of the eclipses at half maximum depth is $0.0236\pm 0.0003$ in orbital phase units.  There is a clear asymmetric ellipsoidal modulation pattern
in which the flux maximum after the deeper eclipses is larger than the maximum
after the shallower eclipses. We attribute this to Doppler beaming, see Section \ref{sec_lc_BF}.

To determine the system
properties, we modelled the light curve with the \texttt{LCURVE} code written
by TRM \citep[for a description of the code, see][Appendix
A]{CopperwheatMarsh2010}. This code uses grids of points to model the two
stars, taking into account limb darkening, gravity darkening, Doppler beaming
and gravitational lensing when the white dwarf eclipses its companion. It assumes the ellipsoidally deformed star to be in corotation with the binary orbit, which is usually a good assumption because of the large tidal interactions between the two binary components. Reprocessing of light from the sdB by the WD is included in the light curve models as well (``reflection effect").

To speed up the
computation of the models used in this paper we implemented a new option 
whereby a finer grid can be used along the track of the white dwarf as it eclipses 
the sdB. This reduces the overall number of points needed to model the 
light curve to the demanding precision required to model the {\em Kepler} data.
In addition, we only used the finely-spaced grid during the eclipse phases,
taking care to make the model values continuous when changing between grids by
applying normalisation factors (very close to unity) to the coarse grid fluxes.
We used $\sim 100\,000$ ($\sim 37\,000$) grid points for the fine (coarse)
grids for the sdB and $3000$ for the white dwarf. To model the finite exposures 
more accurately during the eclipses, where smearing occurs due to the 1\,m integration time, we calculated
7 points for each exposure (i.e., one point for every $\sim 10\,$s) and took their trapezoidal average.


\subsection{Gravity darkening and limb darkening coefficients} \label{sec_lc_gdcldc}
We used model spectra to compute the gravity darkening coefficient (GDC) of
the sdB and the limb darkening coefficients for both the sdB and the white dwarf, which
are all important parameters for the modeling of a close binary's light
curve. The GDC is needed to model the effects of the white dwarf's gravity on the sdB,
which slightly distorts the sdB's shape. The bolometric flux from a stellar
surface depends on the local gravity as $T^4\propto g^{\beta_{b}}$ in which
$\beta_{b}$ is the bolometric GDC. For radiative stars,
$\beta_{b}=1$ \citep{von-Zeipel1924}. We observe the band-limited stellar flux,
not the bolometric flux, and hence we require a different coefficient defined by
$I \propto g^{\beta_{K}}$. The GDC for the {\em Kepler}
bandpass, $\beta_K$ was computed from 
\begin{equation}\beta_{K} = \frac{\mathrm{d} \log
  I}{\mathrm{d} \log g} = \frac{\partial \log I}{\partial \log g} +
\frac{\partial \log I}{\partial \log T}\frac{\mathrm{d} \log T}{\mathrm{d}
  \log g}\end{equation}
in which $I$ is the photon-weighted bandpass-integrated specific
intensity at $\mu=1$ and $\frac{\mathrm{d} \log T}{\mathrm{d} \log
  g}=\frac{\beta_{b}}{4}=0.25$. We used a grid of sdB atmosphere models
calculated from the LTE model atmosphere grid of \citet{HeberReid2000} using the Linfor program \citep{Lemke1997} and assumed
$T_{\rm{eff}}=34\,500$K, $\log g=5.5$, $\log \left(n_{\rm{He}}/n_{\rm{H}}\right) = -1.5$ and
$\log \left(Z/Z_\odot\right)=-2$. To estimate the interstellar reddening, we compared the observed $B-V$ colour of $-0.20\pm0.01\,$mag \citep{AllardWesemael1994} with the colours expected from a model atmosphere. We found an intrinsic colour of $B-V=-0.26\,$mag and consequently adopted a reddening of $E(B-V)=0.06$. To account for this interstellar reddening the model spectra
were reddened following \citet{CardelliClayton1989}. The gravity darkening coefficient was found to be $\beta_K=0.448$.

Using a model for the same set of parameters, we computed limb darkening
coefficients for the sdB. We adopted the 4-parameter limb darkening relation
of \citet[equation 5]{Claret2004LDC} and determined $a_1=0.818$, $a_2=-0.908$,
$a_3=0.755$ and $a_4=-0.252$.

For the white dwarf, angle-dependent model spectra were calculated using the
code of \citet{GansickeBeuermann1995} for $T_{\rm{eff}}=17\,000$\,K (estimated from a comparison of model surface brightnesses given initial light curve fits) and $\log g=7.8$. We adopted the same
limb darkening law as for the sdB and found $a_1=0.832$, $a_2=-0.681$,
$a_3=0.621$ and $a_4=-0.239$. 


\subsection{Doppler beaming factor}\label{sec_lc_BF}
The asymmetry in \target's ellipsoidal modulation pattern is the result of Doppler beaming. Doppler beaming is caused by the stars' radial velocity shifting the
spectrum, modulating the photon emission rate and beaming the
photons somewhat in the direction of motion. The effect was, as far as we are
aware, first discussed in \citet{HillsDale1974} for rotation of white dwarfs and by \citet{ShakuraPostnov1987} for orbital motion in binaries. It was first observed by \cite{MaxtedMarsh2000}. Its expected detection in {\em Kepler} light curves was
suggested and discussed by \citet{LoebGaudi2003} and \citet{ZuckerMazeh2007}.
Van Kerkwijk et al.\ (\citeyear{van-KerkwijkRappaport2010}) report the detection of Doppler beaming in
the long cadence {\em Kepler} light curve of the binary KOI-74. For the first time, they
measured the radial velocity of a binary component from the photometrically
detected beaming effect. The measured radial velocity amplitude, however, did
not match the amplitude as expected from the mass ratio derived from the
ellipsoidal modulation in the light curve.  The derived velocity of the
primary of KOI-74 is yet to be confirmed spectroscopically. For \target,
radial velocities are available which allows the first spectroscopic check of
a photometrically determined radial velocity.

For radial velocities that are much smaller than the speed of light, the
observed flux $F_\lambda$ is related to the emitted flux $F_{0,\lambda}$ as
\begin{equation}F_\lambda = F_{0,\lambda} \left( 1 - B \frac{v_r}{c}\right),\end{equation}
 with $B$ the beaming factor $B = 5+ \rm{d}\ln F_\lambda / \rm{d} \ln \lambda$ \citep{LoebGaudi2003}.
The beaming factor thus depends on the spectrum of the star and the wavelength
of the observations. For the broadband {\em Kepler} photometry, we use a photon
weighted bandpass-integrated beaming factor 
\begin{equation}
\left<B\right> = \frac{\int
  \epsilon_\lambda \lambda F_\lambda B \, d\lambda}{\int \epsilon_\lambda
  \lambda F_\lambda \, d\lambda}
  \end{equation}
   in which $\epsilon_\lambda$ is the response
function of the {\em Kepler} bandpass.

We determined the beaming factor from a series of fully metal line-blanketed LTE models (\citealt{HeberReid2000}, see also Section  \ref{sec_lc_gdcldc}) with metallicities ranging from $\log \left(Z/Z_\odot\right)=-2$ to +1, as well as from NLTE models with zero metals and with Blanchette metal composition (see Section \ref{sec_spec} of this paper for more information about the NLTE models). Without taking reddening into account, the beaming factor is found to be $\left<B\right> = 1.30 \pm 0.03$. The uncertainty incorporates the dependence of the beaming factor on the model grids and the uncertainty on the sdB's effective temperature, gravity and, most importantly, metallicity. The metal composition of the model atmospheres is a poorly known factor that can only be constrained with high-resolution spectroscopy.

This time, the effect of reddening has to be accounted for
by changing the spectral response accordingly instead of reddening the model
atmosphere spectrum. Using a reddened spectrum would in this case erroneously
imply that the reddening is caused by material that is co-moving with the sdB
star. With reddening, the beaming factor is determined to be $\sim 0.006$ lower. Reddening thus only marginally affects the beaming of \target\ but should certainly be taken into account in case of higher reddening values. 

There are three contributions to the beaming factor. The enhanced photon arrival rate of an approaching source contributes $+1$ to the beaming factor. Aberration also increases the number of photons that is observed from an approaching source, adding $+2$ to the beaming factor because of the squared relation between normal angle and solid angle. Finally,  when the sdB comes towards us, an observed wavelength $\lambda_o$ corresponds to an emitted wavelength $\lambda_e = \lambda_o \left(1 + v_r/c\right)$. Since sdBs are blue, looking at a longer 
wavelength reduces the observed flux which counteracts the other beaming 
factor components. In case of an infinite temperature Rayleigh-Jeans spectrum this Doppler shift contribution to the beaming factor would be $-2$. For the primary of \target, we find a contribution of $\sim -1.70$ which brings the total beaming factor to $\sim 1.30$. The contribution of the Doppler shift 
does not always have to be negative; a red spectrum could actually increase 
the effect of beaming.


\subsection{Light curve model} \label{sec_lc_mod}

\begin{figure}
\includegraphics[width=84mm]{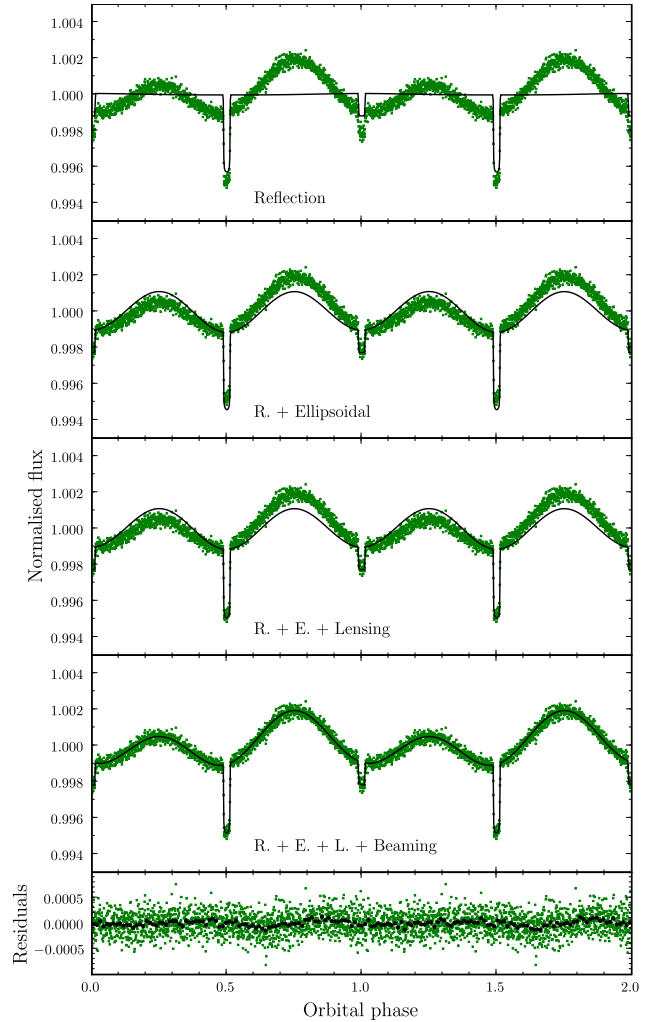}
 \caption{Phase-folded light curve of \target\ (green, datapoints grouped by 30) and our best fitting model (black). In the top panel, only the eclipses and reflection effects are modelled. In the second panel, ellipsoidal modulation is added. In the third panel, gravitational lensing is taken into account as well, which affects the depth of the eclipse at orbital phase 0.5. The bottom panels show the full model -- taking into account Doppler beaming -- and the residuals (grouped by 30 in green and grouped by 600 in black).}
  \label{FIG_LC_mod}
\end{figure}

A typical fit to the data is shown in Fig.~\ref{FIG_LC_mod}, with the different
contributions switched on, one-by-one. From the residuals (bottom panel) it is clear that the model reproduces the variations at the orbital period very well.  When we fit the light curve outside the eclipses with sine curves to represent the reflection effect, ellipsoidal modulation and the beaming, the phase of the ellipsoidal modulation is found to be off by $0.0072 \pm 0.0010$ in orbital phase units. We do not know the origin of this offset, which also gives rise to the shallow structure that is left in the residuals. Our best fits have $\chi^2 = 52032$ for 48929 data points.

The significance of the Doppler beaming is obvious, and even the more subtle gravitational lensing effect is very
significant, although it cannot be independently deduced from the data since
it is highly degenerate with changes in the white dwarf's radius and temperature. One part of the gravitational lensing is caused by light from the sdB that is bent around the white dwarf, effectively making the white dwarf appear smaller. The second lensing contribution is a magnification effect which is caused by the altered area of the sdB that is visible, given that surface brightness is conserved by the lensing effect. In the case of \target, the first part of the lensing is the most important. Lensing effects in compact binaries were discussed in e.g.~\citet{Maeder1973}, \citet{Gould1995}, \citet{Marsh2001} and \citet{Agol2002,Agol2003}.  \citet{SahuGilliland2003} explored the expected influence of microlensing effects on light curves of compact binaries and planetary systems observed by {\em Kepler}. They found that the lensing effect of a typical white dwarf at 1 AU of a main sequence star will swamp the eclipse signal. A transit of a planet, which is of similar size but a lot less massive, can therefore easily be distinguished from an eclipse by a white dwarf. In the case of \target, the separation of the two components is a lot less. An eclipse is still seen, but with reduced depth.
 For the most likely system parameters, gravitational lensing reduces the eclipse depth by $\sim 12$ per cent which is equivalent to a $\sim 6$ per cent reduction of the apparent white dwarf radius. The effect of gravitational lensing is implemented in our light curve modelling code following \citet{Marsh2001}.

\subsection{Markov Chain Monte Carlo simulation} \label{sec_lc_mod}

The parameters which determine models can be fixed by minimisation of
$\chi^2$. If the signal-to-noise is high, a quadratic approximation around the
point of minimum $\chi^2$ can lead to the uncertainties of, and correlations
between, the best-fit parameters. The {\em Kepler} data have superb signal-to-noise,
but owing to the very shallow depths of the eclipses the quadratic approximation does not work well. Strong correlations
between several parameters play a significant role in this problem. The
duration of the eclipses essentially fix the scaled radius of the
sdB star (which we take to be the primary) $r_1 = R_1/a$, where $a$ is the
binary separation. The scaled radius is a function of orbital inclination $i$,
$r_1 = r_1(i)$. The depth of the eclipse of the sdB by the white dwarf fixes the ratio of radii $r_2/r_1 =
R_2/R_1$, so $r_2$ is also a function of orbital inclination. The duration of
the ingress and egress features provides an independent constraint on $r_2$ as
a function of $i$, which can break the degeneracy. In this case, however, one is limited by a combination of signal-to-noise and the
minute-long cadence which is not sufficient to resolve the ingress/egress
features. 

Under these circumstances, a Markov Chain Monte Carlo (MCMC) method
can be very valuable. The MCMC method allows one to build up a sequence of
models in which the fitting parameters, which we denote by the vector
$\mathbf{a}$, have a probability distribution matching the Bayesian posterior
probability of the parameters given the data, $P(\mathbf{a}|\mathbf{d})$. From
long chains of models one can then calculate variances and plot confidence
regions. The MCMC method also has the side benefit of helping with the
minimisation which can become difficult when parameters are highly
correlated. For data in the form of independent Gaussian random variables,
this probability can be written as
\begin{equation} P(\mathbf{a}|\mathbf{d}) \propto P(\mathbf{a}) e^{-\chi^2/2},\end{equation}
i.e., the product of one's prior knowledge of the model parameters and a factor
depending upon the goodness of fit as expressed in $\chi^2$. We implemented
the MCMC method following procedures along the line of
\citet{Collier-CameronWilson2007}. We incorporated prior information in two
ways. In all cases we used our constraint $K_1 = 164 \pm 2\,\kms$. Using our own spectroscopic analysis and the results of \citet{Morales-RuedaMaxted2003}, we decided to put also the following constraint on the effective temperature of the sdB: $T_{1}=34\,500\pm400\,$K. These two constraints were
applied by computing the following modified version of $\chi^2$
\begin{equation} -2 \ln \left(P(\mathbf{a}|\mathbf{d})\right) = \chi^2 + \left(\frac{K_1 -   164}{2}\right)^2+ \left(\frac{T_1 - 34\,500}{400}\right)^2,\end{equation} 
where $K_1$ and $T_1$ are the values in the current MCMC
model under test. The period of the binary orbit was kept fixed at the spectroscopically determined value (see Section \ref{sec_obs}).  The parameters that were kept free during the modelling are the scaled stellar radii $R_1/a$ and $R_2/a$, the mass ratio $q$, the inclination $i$, the effective temperature of the WD $T_2$, the beaming factor and the zero point of the ephemeris. The radial velocity scale (which leads to the masses $M_1$ and $M_2$) and the effective temperature of the primary $T_1$ were included in the fits as well, but with the spectroscopically allowed range as a prior constraint. Note that the beaming factor $\left<B\right>$ is kept as a free parameter, which allows the code to fit the Doppler beaming amplitude while we constrain the allowed range of $K_1$. By comparing the MCMC results for $\left<B\right>$ with the theoretical beaming factor, we can check if the beaming amplitude is consistent with our expectations. 

\begin{figure*}
\includegraphics[width=135mm]{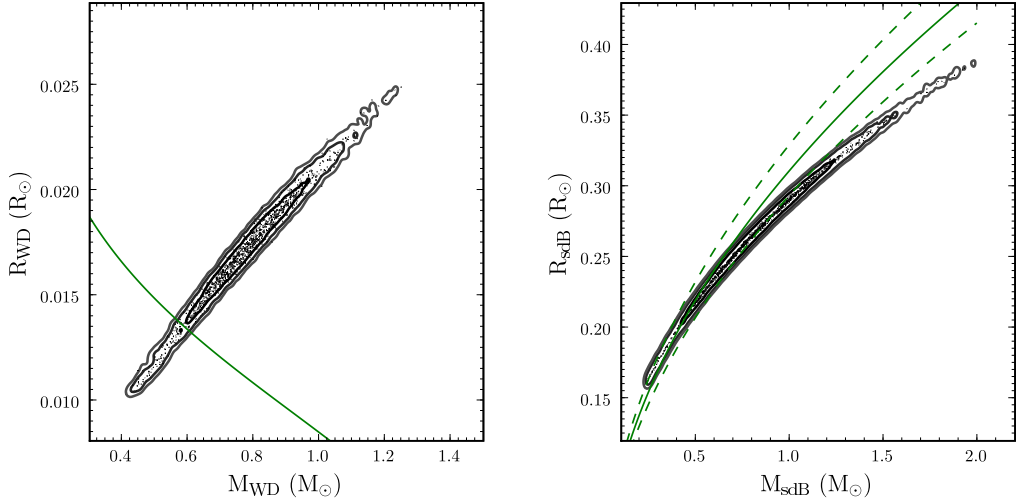}
 \caption{Mass-radius relations of the WD (left) and sdB (right). One tenth of our MCMC models are shown (black dots). The contour plots show the regions in which 68, 95 and 99 per cent of the models reside. The contours that are shown are somewhat artificially broadened by the binning process.  The Eggleton mass-radius relation, inflated by a factor 1.08 to allow for the finite WD temperature (see text for details), is shown as a solid green track on the left plot. The mass-radius relation intersects the Eggleton relation very nearly within its $1$-$\sigma$ region. On the right panel the solid green line gives the mass-radius relation for $\log g=5.45$, the dashed lines for $\log g=5.40$ (left) and $\log g=5.50$ (right). }
  \label{FIG_MR}
\end{figure*}

As explained above, this led to parameter distributions with
strong correlations between $R_1$, $R_2$, $M_1$, $M_2$, $q$, etc. The mass-radius relations for the two stars are shown in Fig.~\ref{FIG_MR}. The
favoured parameters for the secondary (left panel) clearly show that it is a white dwarf
and the allowed distribution nicely crosses the expected track of mass-radius (solid green line)
which we calculated from the zero temperature relation of Eggleton \citep[quoted in][]{VerbuntRappaport1988}, inflated by a factor of $1.08$. We estimated this factor from the cooling models of \citet{HolbergBergeron2006} for white dwarfs with a mass between 0.5 and 0.7\,M$_\odot$ (white dwarfs with masses outside this range are ruled out by the mass-radius relation) using our preferred temperature for the white dwarf of around $16\,000\,$K and assuming an envelope that consists of $M_{\rm{H}} = 10^{-4}\,$M$_{\rm{WD}}$ and 
$M_{\rm{He}}=10^{-2}\,$M$_{\rm{WD}}$. 

\begin{figure*}
\includegraphics[width=135mm]{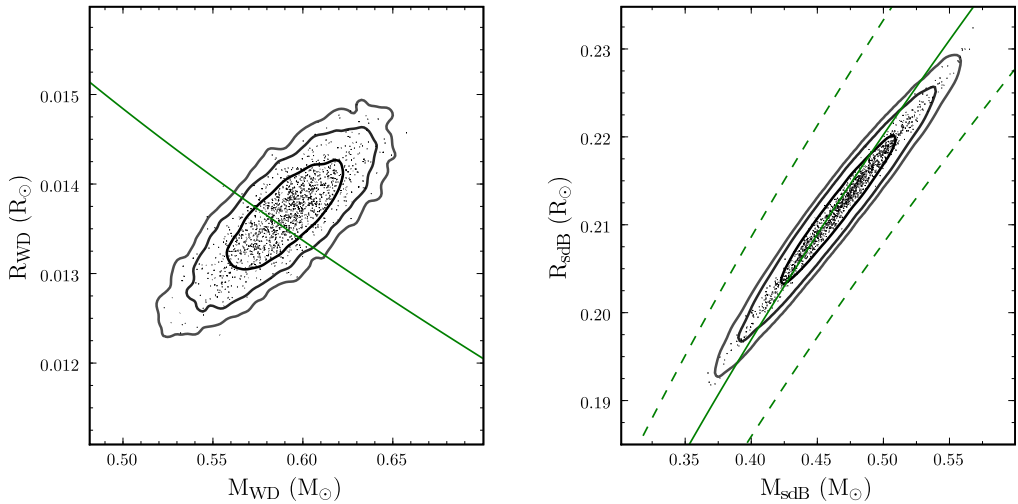}
 \caption{Figure equivalent to Fig.~\ref{FIG_MR} but for an MCMC run with a prior constraint that the white dwarf mass-radius relation has to match the Eggleton relation to within 5 per cent RMS.}
  \label{FIG_MR_WDMR}
\end{figure*}

Given that the secondary is a white dwarf and given that the secondary's theoretical mass-radius relation intersects the mass-radius distribution very nearly within the $1$-$\sigma$ region, we also undertook MCMC runs where
we added the prior constraint that the secondary had to match the white dwarf
M-R relation to within an RMS of 5 per cent. This was added in exactly the same
manner as the $K_1$ and $T_1$ constraints. The system parameters we derive from these MCMC runs are listed in Table \ref{tab_pars}. The mass-radius relation after applying the constraint is shown in Fig.~\ref{FIG_MR_WDMR}. Especially after applying the white dwarf mass-radius relation constraint, the sdB's mass-radius relation fits perfectly with the one defined by the surface gravity derived from spectroscopy in Section \ref{sec_spec}. The correlation coefficients between the different parameters are given in Table \ref{tab_corr}. The binary's inclination, its mass ratio and the stellar radii are highly correlated. 

\begin{table}
 \caption{Properties of \target. The orbital period and the effective temperature of the sdB were derived from spectroscopy. The other parameters are obtained by modelling the {\em Kepler} light curve. The uncertainties on these values are determined by MCMC analysis, using the prior constraint that the white dwarf mass-radius relation has to match the Eggleton relation to within 5 per cent RMS. }
 \label{tab_pars}
 \begin{center}
 \begin{tabular}{lcc}
  \hline
  & Primary (sdB) & Secondary (WD) \\
  \hline
$P_{\rm{orb}}$ (d) &  \multicolumn{2}{c}{$0.40375026(16)$} \\
$q$ & \multicolumn{2}{c}{$1.27 \pm 0.06$} \\
$i$  (deg) & \multicolumn{2}{c}{$87.14 \pm 0.15$} \\
$R$ $\left(\rm{R}_\odot\right)$&  $0.212 \pm 0.006$ & $0.0137 \pm 0.0004$ \\
$M$ $\left(\rm{M}_\odot\right)$&  $0.47 \pm 0.03$ & $0.59 \pm 0.02$ \\
$T_{\rm{eff}}$ (K) &  $34$\,$500\pm400$ & $15$\,$900 \pm 300$\\
\hline

  \end{tabular} \end{center}
\end{table}

\begin{table}
 \caption{Correlation coefficients of the different parameters that were varied in the MCMC simulations, after applying the Eggleton mass-radius relation constraint. }
 \label{tab_corr}
 \begin{center}
 \begin{tabular}{lccccc}
  \hline
             & $R_2$ & $i$     &  $T_1$ & $T_2$ & $q$ \\
             \hline
 $R_1$ &  0.95    & -0.95 &   0.02    &  0.02   & -0.95 \\
 $R_2$ &             & -0.98 &   0.07    &  0.02   & -0.99 \\
 $i$       &             &          &   -0.06   & -0.02  &  0.96 \\
 $T_1$ &             &           &              &  0.44  & -0.09 \\ 
 $T_2$ &             &           &             &            &  -0.02 \\
 
 \hline

  \end{tabular} \end{center}
\end{table}

\subsection{Variability in residuals}
One of the goals of the {\em Kepler Mission} is to allow detailed asteroseismic studies of pulsating stars. The asteroseismology programme is discussed in \citet{GillilandBrown2010}.  For more information about the search for pulsations in compact objects with {\em Kepler}, see \citet{OstensenSilvotti2010}. 

The Fourier transform of the original light curve of an eclipsing close binary like \target\ is highly contaminated by frequencies and their harmonics due to the binary orbit. Subtraction of a good model of the binary signatures of the light curve allows one to get rid of this contamination. Since a number of sdBs have been found to be multiperiodically pulsating \citep[for a review on asteroseismology of EHB stars, see][]{Ostensen2009}, we checked the residuals of the light curve for signs of pulsations. Using the analysis method and significance criteria outlined in \citet{DegrooteAerts2009}, 8 significant frequencies were found, which are listed in Table~\ref{tab_freqs}. 

\begin{table}
 \caption{Significant variability frequencies in the residuals of the light curve of \target. The signal to noise (S/N) value was determined by dividing the amplitude of the peak by the uncertainty on the amplitude.}
 \label{tab_freqs}
 \begin{center}
 \begin{tabular}{lrcrl}
  \hline
  & Frequency (d$^{-1}$) & Amplitude& S/N & \\
  & &  ($\mu$mag) & &\\
  \hline
$f_1$& $    0.2758  \pm 0.0011$&  $ 97.5\pm  7.2$&$     13.5  $ & $=f_2/2$\\
$f_2$& $    0.5936  \pm0.0013 $&  $ 97.5\pm  7.2$&$    13.5   $ & instrumental\\
$f_3$& $    0.1417  \pm0.0014 $&  $ 86.6\pm  7.2$&$     12.1  $& $=f_2/4$\\
$f_4$& $    1.1820  \pm0.0015 $&  $ 81.0\pm  7.2$&$     11.3  $& $=2f_2$\\
$f_5$& $    1.7730  \pm0.0018 $&  $ 67.5\pm  7.2$&$     9.4  $& $=3f_2$\\
$f_6$& $  440.4386 \pm 0.0022$& $ 54.2\pm  7.2$& $     7.6 $ & instrumental\\
$f_7$& $    4.9548  \pm0.0024  $& $ 49.1\pm  7.2$& $     6.9 $ & $=2f_{orb}$\\
$f_8$& $    0.3115  \pm0.0027 $&  $ 44.9\pm  7.2$&$     6.3  $& \\
\hline
  \end{tabular} \end{center}
\end{table}

$f_2$ is a known artefact frequency caused by an eclipsing binary that was used as one of the fine-guidance stars during Q1 \citep[see][]{HaasBatalha2010, JenkinsCaldwell2010}. Four other frequencies ($f_1$, $f_3$, $f_4$ and $f_5$) are related to $f_2$. 
The highest frequency, $f_6$, is related to the processing of the long cadence data \citep[see][]{GillilandJenkins2010}. $f_7$ is the first harmonic of the orbital frequency of \target, which indicates that there is still a weak orbital component left after subtracting our light curve model. $f_8$ is not related to any of the other frequencies and corresponds to a period that is too long to arise from stellar pulsations of the WD or the sdB. If it is real, the signal might result from the rotation of the WD, or it might be due to a background star.

The best candidate peak for $p-$mode pulsations of the sdB is at $5018.2\,\mu$Hz with an amplitude of $37\, \mu$mag, but further data is needed to confirm that the sdB is pulsating. From ground based data, \citet{OstensenOreiro2010} did not detect pulsations, with a limit of 0.68\,mma. This is consistent with the Kepler photometry.


\section[]{Spectroscopic analysis} \label{sec_spec} 
Low resolution high S/N spectra for \target\ were taken with the
B\&C spectrograph at Steward Observatory's 2.3-m Bok telescope on Kitt
Peak, as part of a long term homogeneous survey of hot subdwarf stars
\citep{GreenFontaine2008}, in September and October 2004.  The spectrograph parameters,
observational procedures, and reduction techniques were kept the same
for all the observing nights.

The 400/mm grating, blazed at $4889\,\rm\AA$, gives a resolution of R = 560
over the wavelength region $3620-6895\,\rm\AA$, when used with the 2.5 arcsec slit.
The spectra were taken during clear or mostly clear conditions with integration times between 1050 and 1200\,s.
Approximately 1000 bias and flat-field images were obtained for
each run.  The data reductions were performed using standard IRAF
tasks, and each night was flux-calibrated separately.  The individual
spectra were cross-correlated against a super-template to determine
the relative velocity shifts, and then shifted and combined into a
single spectrum.  Although the resolution is rather low, the S/N is
quite high: 221/pixel or 795/resolution element for the combined
spectrum.  The continuum fit to the combined flux-calibrated spectrum
were done with great care to select regions devoid of any weak lines,
including expected unresolved lines of heavier elements.

The final \target\ spectrum was fitted using two separate grids of
NLTE models designed for sdB stars, in order to derive the effective
temperature, surface gravity and He/H ratio.  The first set of models
assumed zero metals, while the second included an adopted distribution
of metals based on the analysis of FUSE spectra of five
sdB stars by \citet{BlanchetteChayer2008}, see also \citet{Van-GrootelCharpinet2010}. From the set of models without metals, we derive 
$\log g=5.45\pm0.04$, $T_{\rm{eff}}=34\,400\pm220\,$K and $\log(\rm{He}/\rm{H})=-1.37\pm0.05$. Assuming the Blanchette composition, we find 
$\log g=5.43\pm0.04$, $T_{\rm{eff}}=34\,730\pm250\,$K and $\log(\rm{He}/\rm{H})=-1.36\pm0.04$. These results are in good agreement with 
$\log g=5.37\pm0.10$, $T_{\rm{eff}}=34\,500\pm1000\,$K, $\log(\rm{He}/\rm{H})=-1.35\pm0.10$ 
determined by \citet{Morales-RuedaMaxted2003} and $\log g=5.43\pm0.10$, $T_{\rm{eff}}=34\,200\pm500\,$K  
determined by \citet{GeierHeber2010} using different spectra and model grids.

\begin{figure*}
\includegraphics[width=110mm]{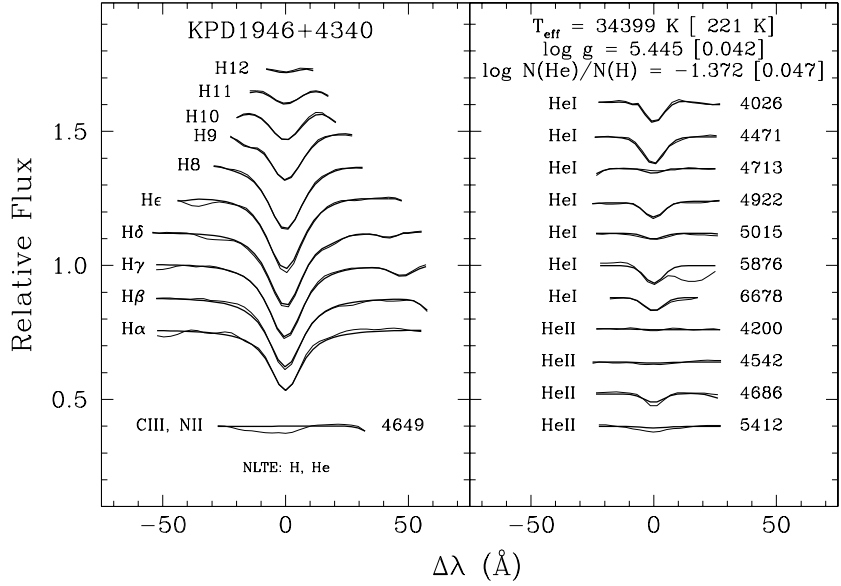}
 \caption{Fit (bold lines) to the spectral lines of \target\ using NLTE sdB models assuming zero metals. }
  \label{FIG_Fnometals}
\end{figure*}

\begin{figure*}
\includegraphics[width=110mm]{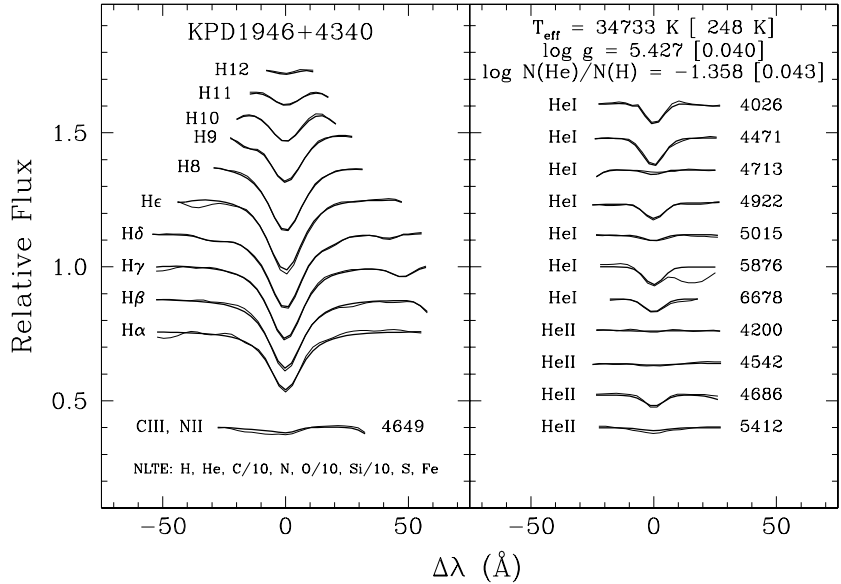}
 \caption{Fit (bold lines) to the spectral lines of \target\ using NLTE sdB models assuming a Blanchette metal composition.}
  \label{FIG_Fmetals}
\end{figure*}

The fit definitely improves when going from the zero-metal solution (Fig.~\ref{FIG_Fnometals}) to
the Blanchette composition (Fig.~\ref{FIG_Fmetals}), although there still remains a slight ``Balmer'' problem, especially noticeable in the core of H$\beta$. There
are definitely metals in the spectrum of \target: the strongest
features are 1) an unresolved C\,{\textsc III} + N\,{\textsc II} complex around $4649\,\rm\AA$
(compare the two figures for that feature), and 2) another weaker
complex (C\,{\textsc III} + O\,{\textsc II}) in the blue wing of H$\delta$ that the Blanchette
model reproduces quite well.  All of the major discrepancies between
the spectra and the models are due to strong interstellar absorption:
the K line of Ca\,{\textsc II} in the blue wing of H$\epsilon$, the Ca\,{\textsc II} H line in the
core of H$\epsilon$, and the Na\,{\textsc I} doublet strongly affecting the red wing
of He\,{\textsc I} 5876.  It is reassuring that the derived atmospheric
parameters are not too strongly dependent on the presence of metals,
as might be expected for such a hot star, particularly one in which
downwards diffusion of metals is important.

\section{Discussion}\label{sec_disc}
\begin{figure}
\includegraphics[width=84mm]{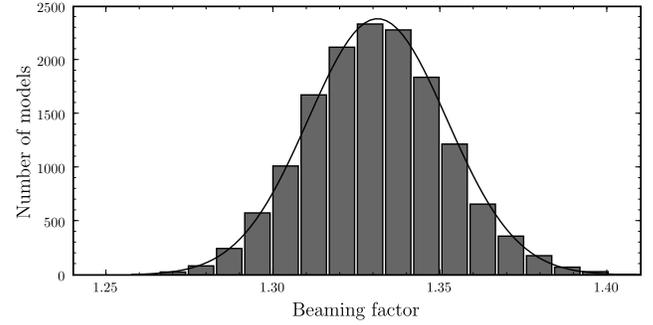}
 \caption{Distribution of the sdB's beaming factor for an MCMC run (using the M-R constraint for the WD; see text for details). The beaming factor is found to be $\left<B\right> = 1.33\pm0.02$, which is in agreement with the theoretically expected $\left<B\right> = 1.30 \pm 0.03$.} 
  \label{FIG_BF}
\end{figure}

The beaming factor we derived for \target\ using MCMC runs is $\left<B\right> = 1.33\pm0.02$, which is in perfect agreement with the theoretically expected value calculated in Section \ref{sec_lc_BF}. The uncertainty on the beaming factor is a direct reflection of the uncertainty on the spectroscopic radial velocity amplitude of the sdB. If, contrary to our assumption, the {\em Kepler} fluxes would be severely contaminated by light from other (constant) stars, the observed beaming factor would be lower. The distribution of beaming factors from our MCMC computations is shown in Fig.~\ref{FIG_BF}. 
If the radial velocity would be measured from the Doppler beaming amplitude, using the theoretical beaming factor, the result would be $168\pm 4\,\kms$ compared to $164.0\pm1.9\,\kms$ derived from spectroscopy. The uncertainty on the photometric radial velocity is dominated by the uncertainty on the theoretical beaming factor, primarily due to its dependence on the poorly known metallicity of the sdB, and to a lesser extent due to the uncertainties on the sdB's effective temperature and surface gravity.

Under the assumption of corotation, we find a projected rotational velocity of the sdB of $v \sin(i) = 26.6 \pm 0.8\, \kms$.  From spectroscopy and using LTE models with ten times Solar metallicity, \citet{GeierHeber2010} found $v \sin(i) = 26.0 \pm 1.0\, \kms$, which is in agreement with our photometric result. We conclude that the assumption of corotation is likely to be correct. 

The spectroscopically determined surface gravity of the sdB ($\log g=5.43\pm0.04$ and $5.45\pm0.04$ using atmosphere models with and without metals respectively) agrees perfectly with the surface gravity of $5.452\pm0.006$ we derived from the mass-radius distribution of our light curve models.

\begin{figure}
\includegraphics[width=84mm]{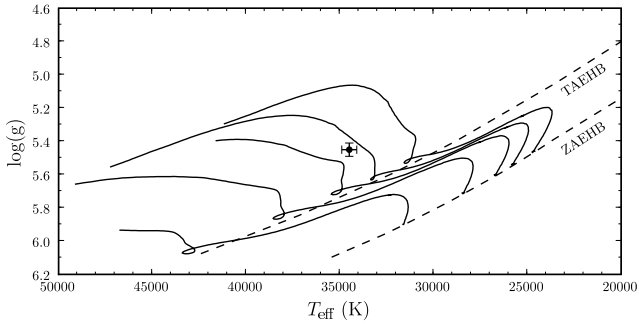}
 \caption{The sdB of \target\ in the $T_{\rm{eff}}-\log g$ plane. The theoretical zero age and terminal age extended horizontal branches for a $0.47\,$M$_\odot$ are shown, together with evolutionary tracks for different envelope thicknesses ($10^{-4}$, $10^{-3}$, $2\times10^{-3}$, $3\times10^{-3}$ and $4\times10^{-3}\,$M$_\odot$). The sdB is found to be in the post-EHB phase.}
  \label{FIG_evolution}
\end{figure}

As concluded earlier by \citet{Morales-RuedaMaxted2003}, the sdB is probably in a post-EHB phase. This is illustrated in Fig.~\ref{FIG_evolution}, which shows the zero age extended horizontal branch (ZAEHB) and the terminal age extended horizontal branch (TAEHB) for an sdB with a typical core mass of $0.47\,$M$_\odot$, together with evolutionary tracks for different hydrogen envelope masses ($10^{-4}$, $10^{-3}$, $2\times10^{-3}$, $3\times10^{-3}$ and $4\times10^{-3}\,$M$_\odot$) from \citet{KawalerHostler2005}. 

Because of its low surface gravity, the sdB component of KPD1946 falls in
a region of the $T_{\rm{eff}}-\log g$ plane relatively far from the center of the
instability strip. However, at least one pulsator exists in this region of the $T_{\rm{eff}}-\log g$
plane, V338 Ser, that should be in a post-EHB phase \citep[see][Fig.~3]{Ostensen2009}.
Moreover, `transient pulsators' with varying pulsation amplitudes that
can go down to undetectable values in a particular epoch might exist
\citep[see the case of KIC 2991276 in][]{OstensenSilvotti2010}.
For these reasons, and because we found at least one candidate $p-$mode pulsation frequency, it is worth continuing a photometric monitoring by Kepler.

The white dwarf mass implies that it is a CO white dwarf. The progenitor of the sdB must have been the less massive star in the original binary and by the time it reached the ZAEHB, the white dwarf was already cooling. The accretion of material by the white dwarf does not change the WD's internal energy content significantly (see e.g.~related work on cataclysmic variables by \citealt{TownsleyBildsten2002}). The cooling time of the white dwarf
therefore sets an upper limit to the time since the sdB was on the ZAEHB. For our best
estimates of the temperature and mass of the white dwarf, the cooling tracks of \citet{HolbergBergeron2006} indicate that it has been cooling for about $155$ to $170$\,Myr (depending on the unknown envelope composition). The sdB's evolution from the Zero Age Extended Horizontal Branch to its current post-EHB-phase took 125 to 145\,Myr \citep[][depending on the exact current evolutionary stage]{KawalerHostler2005}, which means that the sdB must have formed very shortly after the white dwarf.

\section[]{Summary} \label{sec_concl}
We have analysed a $33.5$-d short cadence {\em Kepler} light curve of \target, as well as low resolution spectroscopy. In the light curve, primary and seconday eclipses, ellipsoidal modulation and Doppler beaming are detected. We model the binary light curve, taking into account the Doppler beaming and gravitational lensing effects. System parameters and uncertainties are determined using Markov Chain Monte Carlo simulations. 

The binary is found to consist of a $0.59\pm0.02\,$M$_\odot$ white dwarf and a $0.47\pm0.03\,$M$_\odot$ post-EHB sdB star. The surface gravity and corotation rotational velocity of the sdB as derived from the light curve models are found to be consistent with spectroscopic values. The observed Doppler beaming amplitude is in perfect agreement with the amplitude expected from spectroscopic radial velocity measurements. It would thus have been possible to derive the radial velocity amplitude of the sdB from the Kepler light curve directly. 

Subtracting a good light curve model allowed us to search for stellar oscillations. No significant stellar variability of the sdB or white dwarf could be detected yet. At least one candidate $p-$mode pulsation frequency was found, however, and the sdB can also possibly be a transient pulsator. \target\ continues to be observed by {\em Kepler}.


\subsection*{Acknowledgments}
We thank the referee Martin van Kerkwijk for his helpful suggestions.
The authors gratefully acknowledge everybody who has contributed to make the {\em Kepler Mission}  possible. Funding for the {\em Kepler Mission} is provided by NASA's Science Mission Directorate.
Part of the data presented here have been taken using ALFOSC, which is owned by the Instituto de Astrofisica de Andalucia (IAA) and operated at the Nordic Optical Telescope (Observatorio del Roque de los Muchachos, La Palma) under agreement between IAA and the NBIfAFG of the Astronomical Observatory of Copenhagen. This research also made use of data taken with the Bok telescope (Steward Observatory, Kitt Peak).
The research leading to these results has received funding from the European
Research Council under the European Community's Seventh Framework Programme
(FP7/2007--2013)/ERC grant agreement n$^\circ$227224 (PROSPERITY), as well as
from the Research Council of K.U.Leuven grant agreement GOA/2008/04.  During
this research TRM was supported under grants from the UK's Science and
Technology Facilities Council (STFC, ST/F002599/1 and PP/D005914/1).


\bibliographystyle{mn}
\bibliography{KPD1946.bib}

\label{lastpage}
\end{document}